\documentclass{article}
\usepackage{spconf,graphicx}
\usepackage{url}
\usepackage{multirow}
\usepackage{enumitem}
\usepackage{xcolor} 
\usepackage{latexsym}
\usepackage[linesnumbered,ruled,vlined]{algorithm2e}
\usepackage{amsmath}

\title{Real-time, Universal, and Robust Adversarial Attacks \\Against Speaker Recognition Systems\vspace{-2mm}}
\name{ Yi Xie$^1$, Cong Shi$^1$, Zhuohang Li$^2$, Jian Liu$^2$, Yingying Chen$^1$, Bo Yuan$^1$\vspace{-2mm}}

\address{$^1$Rutgers University, New Brunswick, NJ, USA, 08901\\
$^2$The University of Tennessee, Knoxville, TN, USA, 37996\vspace{-3mm}}
\begin{document}
%
\maketitle
\vspace{-4mm}
\begin{abstract}
\vspace{-1mm}
As the popularity of voice user interface (VUI) exploded in recent years, speaker recognition system has emerged as an important medium of identifying a speaker in many security-required applications and services. In this paper, we propose the first real-time, universal, and robust adversarial attack against the state-of-the-art deep neural network (DNN) based speaker recognition system. 
Through adding an audio-agnostic universal perturbation on arbitrary enrolled speaker's voice input, the DNN-based speaker recognition system would identify the speaker as any target (i.e., adversary-desired) speaker label. 
In addition, we improve the robustness of our attack by modeling the sound distortions caused by the physical over-the-air propagation through estimating room impulse response (RIR). Experiment using a public dataset of $109$ English speakers demonstrates the effectiveness and robustness of our proposed attack with a high attack success rate of over $90\%$. The attack launching time also achieves a $100\times$ speedup over contemporary non-universal attacks.

\end{abstract}
\vspace{-2mm}
\begin{keywords}
speaker recognition systems, adversarial examples, universal adversarial attack
\end{keywords}

\vspace{-4mm}
\section{Introduction}
\label{sec:intro}
\vspace{-2.5mm}
In recent years, voice user interface (VUI) has been integrated into various platforms, such as smartphones and smart appliances, and is shaping up to become the hubs of our increasingly connected lives. 
With the prevalent usage of VUI, speaker recognition system, which identifies a person from characteristics of voices, could be seamlessly integrated and used for various security-enhanced applications, such as remote voice authentication to prevent fraud in financial services, voice-matched voice assistants that can only respond to the owner's voice, and even suspects identification and criminals detection~\cite{voice_match,chase}.

Deep network networks (DNNs), with its superiority over current state-of-the-art models (e.g., universal background model-Gaussian mixture model)~\cite{lei2014novel,mclaren2015advances}, has been becoming the computation core of the speaker recognition systems. However, recent studies have shown that DNN models are vulnerable to adversarial input in various fields (e.g., computer vision~\cite{intriguing}, natural language processing~\cite{carlini2018audio,carlini2016hidden} and speaker verification~\cite{kreuk2018fooling}). The most related work~\cite{kreuk2018fooling} generates adversarial examples against an end-to-end speaker verification model, which is a binary speaker recognition system that verifies whether the voice is uttered by a claimed speaker or not. However, the adversarial attack against a more complex multi-class speaker recognition model still remains unexplored. Moreover, this attack~\cite{kreuk2018fooling} is \textit{individual attack} (i.e., non-universal) requiring to generate different perturbation for each voice input, which would cost considerable time training perturbations for each individual voice input and thus make real-time attacks impossible. 

In this paper, we build the first \textit{real-time}, \textit{universal}, and \textit{robust} targeted adversarial attack on X-vector~\cite{snyder2018x}, a state-of-the-art DNN-based multi-class speaker recognition model.
The adversarial attack is performed by crafting an audio-agnostic universal perturbation which can be added into any enrolled speaker's any voice input to deceive the speaker recognition system, causing it to output an adversary-desired (targeted) speaker label. The generated universal perturbation uses repeated-playback of fixed-length universal noise to fit different voice input with various lengths.
Additionally, unlike the existing digital attack~\cite{kreuk2018fooling} that feeds the adversarial examples to the speaker verification model directly, in this paper we take one step forward to build robust adversarial attacks through estimating the sound distortions introduced by the physical world propagation, which makes the adversarial examples remain effective while being played over-the-air. Experiments on a public dataset of $109$ speakers show the effectiveness and robustness of our proposed attack with a high attack success rate of over $90\%$. The achieved attack launching time is only around $0.015s$, which is $100\times$ speedup over contemporary non-universal attacks.

\vspace{-5mm}
\section{Related Work}
\label{sec:related}
\vspace{-2.5mm}

\textbf{Adversarial Attack on Speech Recognition.}
Recent studies have successfully produced adversarial examples against automatic speech recognition (ASR) system (i.e., speech-to-text), which is the most prevalent application in the audio space. For instance, Vaidya \textit{et al.}~\cite{vaidya2015cocaine,carlini2016hidden} generate noise-like adversarial sound making ASR models output adversary-desired text transcriptions.
Nonetheless, the generated adversarial examples would be perceived as noises by human, which may draw considerable attention on practical attacks. To solve this problem, Carlini \textit{et al.}~\cite{carlini2018audio} propose to craft adversarial samples by adding unnoticeable perturbations into original speech, misleading the model to translate the adversarial examples to adversary-desired text. Moreover, CommanderSong~\cite{yuan2018commandersong} can embed any malicious command into regular songs, which could be recognized by ASR systems as malicious commands but still being perceived as common music by human. However, all the aforementioned ASR adversarial attacks are individual attack through solving an optimization problem for each individual input audio, which needs high run-time requirements (e.g., several hours) to compute the adversarial examples per input audio.
Alternatively, a more recent work~\cite{neekhara2019universal} produces a single universal perturbation which can fool ASR systems causing an error in transcription. This work is in the case of untargeted attack, in which the adversary cannot specify the expected speech transcription during the phase of adversary example generation.

\noindent\textbf{Adversarial Attack on Speaker Recognition.}
Different from speech recognition systems, speaker recognition (a.k.a., voice recognition) mainly focuses on extracting individual-dependent voice characteristics through embedding methods to identify speakers' identities regardless of their speech content. It has been shown a growing trend of using DNNs in the embedding layers of speaker recognition model due to its superiority of scalable embedding performance~\cite{lei2014novel,mclaren2015advances}. However, few studies have been conducted to explore the vulnerability of the DNN-based speaker recognition system. To the best of our knowledge, the only related study~\cite{kreuk2018fooling} proposes to build adversarial examples against an end-to-end speaker verification model, which is a binary speaker recognition system. Moreover, this attack is \textit{individual attack}, which requires a long time to craft different perturbation for each voice input. It does not consider any sound distortions caused by practical over-the-air playback either. To bridge the gap in terms of all the aforementioned issues, in this paper we explore the possibility of launching real-time universal, targeted, and robust adversarial attacks against multi-class speaker recognition system, with $109$ speakers in our testing model.

\vspace{-4mm}
\section{Real-time, Universal, and Robust Adversarial Examples}
\label{sec:universal}
\vspace{-3mm}

\subsection{Target Speaker Recognition Model}
\vspace{-2mm}

\begin{figure}[t]
\centering
\includegraphics[width=0.39\textwidth]{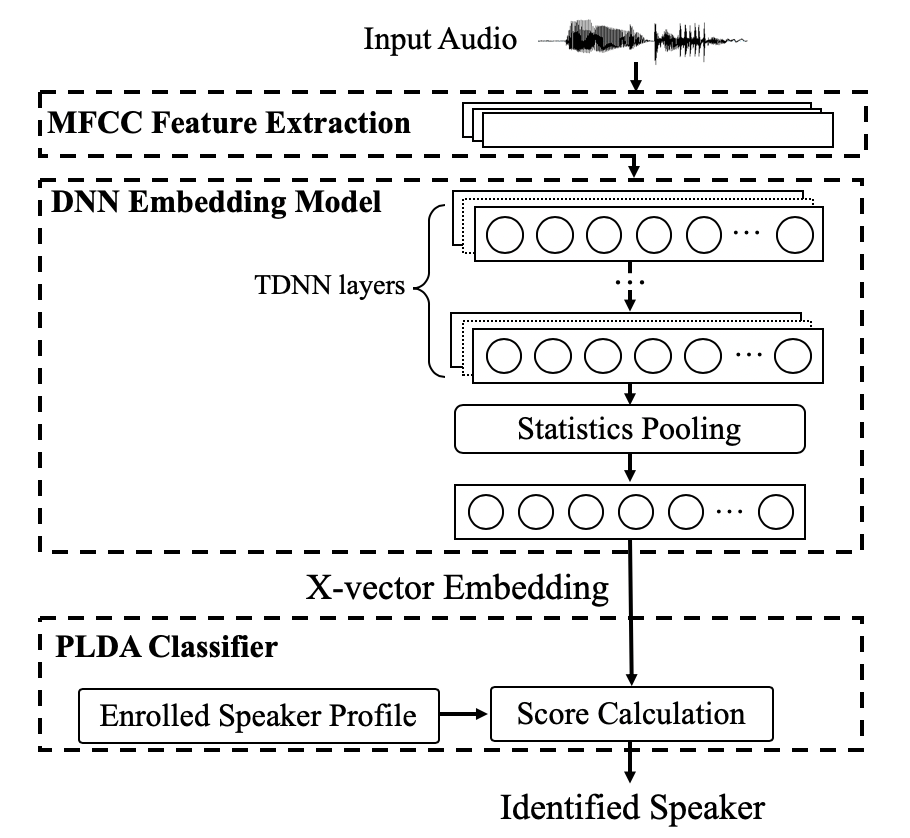}
\vspace{-4mm}
\caption{Targeted speaker recognition model (X-vector).}
\vspace{-4mm}
\label{fig:xvector}
\vspace{-2mm}
\end{figure} 
In this work, the DNN-embedding-based X-vector system~\cite{snyder2018x} is used as the speaker recognition system since it has shown a significant improvement over standard i-vector models, and has been further studied in many follow-up studies (e.g.,~\cite{snyder2019speaker,raj2019probing}).
The architecture of X-vector system is shown in Figure~\ref{fig:xvector}. Specifically, for an input audio, the system first extracts mel-frequency cepstral coefficents (MFCCs) features using a sliding window. The extracted features are then passed to a time-delay neural network (TDNN) structure~\cite{snyder2017deep} that operates on audio frames. The statistics pooling layer takes the output of the final frame-level layer as input, aggregates over the input segment, and computes its mean and standard deviation.
Subsequently, hidden layers are used to map the concatenated statistics into final embeddings. In the recognition phase, the probabilistic linear discriminant analysis (PLDA) computes the probability of the input audio belonging to each enrolled speaker with the embedding information and identifies the speaker label with the highest calculated score.

\vspace{-5mm}
\subsection{Challenges and Threat Model}
\vspace{-2mm}
\textbf{Challenges.}
Generating such a real-time, universal, and robust adversarial example against speaker recognition system in practice raises a number of challenges:

\noindent \textit{(1) Real-time Adversarial Attack.} To craft an adversarial noise with respect to the speaker's speech, using conventional optimization-based approach is usually very time-consuming, which makes many practical attack scenarios impossible, such as playing the adversarial noise on a hidden speaker in a real-time manner along with the speaker's voice input.

\noindent \textit{(2) Universal Targeted Adversarial Example.} Using an audio-agnostic universal perturbation to deceive the speaker recognition system, which causes it to misclassify any enrolled speaker's input audio as the adversary-desired speaker, needs to build a universal mapping from the audio sources to the adversary-desired target. The proposed algorithm needs to be general enough to various length audio inputs spoken by different speakers with various accents.

\noindent \textit{(3) Robust Adversarial Example.} The attack performance would be inevitably impacted by the sound distortions due to the attenuation and multi-path effects while playing the adversarial examples over the air. Thus, the generated adversarial perturbation needs to be robust enough to remain effective under this kind of real-world distortions.

\begin{figure}[t]
\centering
\includegraphics[width=0.45\textwidth]{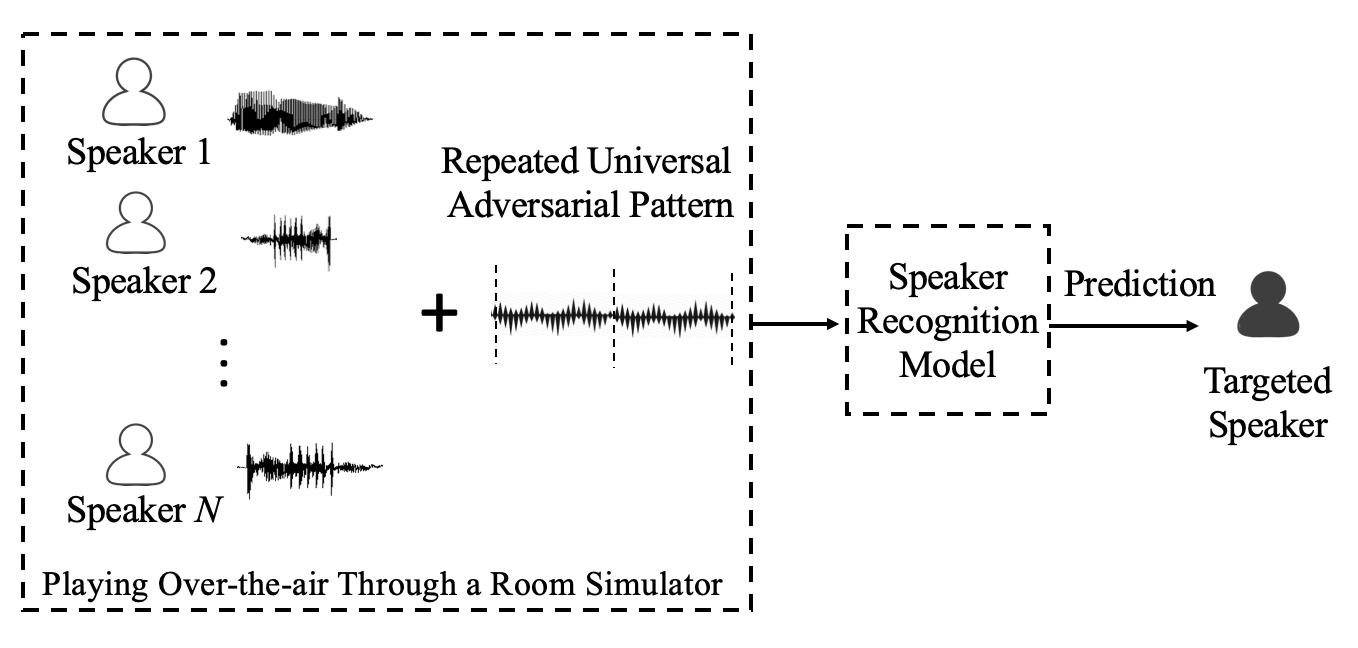}
\vspace{-5mm}
\caption{Threat model of the proposed attack.}
\vspace{-4mm}
\label{fig:attack_model}
\vspace{-2mm}
\end{figure}

\noindent \textbf{Threat Model.}
In this work, we consider the white box threat model where the adversary has full knowledge of the target speaker recognition model as well as its parameters. In order to build a robust adversarial attack considering the sound distortions in the room where the attack will be launched, we assume the adversary has access to the room's layout.
As shown in Figure~\ref{fig:attack_model}, we aim to find a single audio-agnostic universal perturbation that can be applied on arbitrary enrolled speakers' input audio to mislead the speaker recognition system causing it output the specific adversary-desired speaker label. Additionally, we expect to build a more robust adversarial perturbation that can remain effective while being played over-the-air in acoustic room simulated environments.

\vspace{-5mm}
\subsection{Real-time, Universal and Robust Adversarial Attacks}
\label{3-3}
\vspace{-2mm}

Most of the existing targeted adversarial attacks would fool DNN-based systems through building different adversary perturbation for each individual input. Differently, in this paper we explore how to build a single universal perturbation that can be directly applied to arbitrary speaker's any utterance, making the speaker recognition system output the adversary-desired speaker label. Such a universal perturbation would greatly shorten the attack launching time, making real-time attacks possible.

To clearly present the steps of our perturbation generation, we model the target speaker recognition system, X-vector, as a function $F(x)$, which takes as input an utterance $x$ and outputs a predicted speaker label. We define $P(x)$ as the function of all DNN layers (including PLDA) to compute the probabilities  of  classifying $x$ as each of the profiled speakers.
We can recognize the voice as the speaker with highest calculated probability, $F(x)=Argmax(P(x))$. 
Therefore, to launch a universal targeted adversarial attack, where targeted speaker label is $t$, we aim to find a perturbation $\delta$ that could achieve $F(x+\delta)=Argmax(P(x+\delta))=t$ for arbitrary $x$.

To build such a universal attack, we need to find a general solution that can make the generated perturbation effective for all the utterances regardless of their speakers, accents, speech content and length. To overcome the issue of varying utterance length, we dynamically construct the universal perturbation $\delta$ based on the length of the input utterance $x$:
\begin{equation}
\setlength{\abovedisplayskip}{3pt}
\setlength{\belowdisplayskip}{3pt}
\begin{aligned}
\delta = Crop([\bigtriangleup \delta\frown...\frown\bigtriangleup \delta], x),
\end{aligned}
\label{eq:crop}
\end{equation}
where $\bigtriangleup \delta$ is a short-length adversarial perturbation (e.g., $1s$ in our work), and $[\bigtriangleup \delta\frown...\frown\bigtriangleup \delta]$ is a vector constructed by repeating $\bigtriangleup \delta$. $Crop(\cdot, \cdot)$ crops the first input to the length of the second input. With this process, the derived perturbation $\delta$ could be applied to the audio input with any length.

To minimize the distortion between the adversarial example and the original voice, $\delta$ would be clipped to a pre-defined range. The generated adversarial example with the clipped $\delta$ could be formulated as: 
\begin{equation}
\setlength{\abovedisplayskip}{3pt}
\setlength{\belowdisplayskip}{3pt}
    x' = x+Clip_{\epsilon}(\delta),
\label{eq:adv_expl}
\end{equation} 
where $Clip_{\epsilon}(\delta)$ is the function to perform element-wise clipping of $\delta$. Values of $\delta$ outside the interval $[-\epsilon, \epsilon]$ would be clipped to the interval edges, and $\epsilon$ is our pre-defined attack strength.

Moreover, to preserve the effectiveness of the adversarial example while being played over the air, we first mimic the sound distortions during playback and recording by estimating room impulse response (RIR), $r$, which characterizes the acoustic propagation (e.g., reverberations) in a room environment.  
The details of how to estimate RIR (i.e., $r$) based on the room setting are provided in Section~\ref{sec:rir}. Then, we could iteratively derive the targeted adversarial example through the following objective function:  
\begin{equation}
\setlength{\abovedisplayskip}{3pt}
\setlength{\belowdisplayskip}{3pt}
Argmax(P(x'*r)) = t, 
\end{equation}
where $t$ is the targeted speaker label, $*$ denotes the convolution operation, and $x'*r$ is the estimated adversarial example recorded by the microphone. 
It is important to note that the estimated RIR represents a certain mapping from the played sound to the recorded sound as per specific location of the loudspeaker and microphone in the room. To make the generated adversarial examples robust in various environmental settings, we estimate multiple RIRs $\mathbf{r}$ in various environments. To make the adversarial perturbation survive all these environments, we randomly select one RIR in $\mathbf{r}$ for each training step when updating the perturbation based on each training utterance.
In addition, as directly solving the non-linear constrained non-convex problem is difficult, we iteratively solve the following optimization problem\cite{carlini2016hidden}:  
\begin{equation}
\setlength{\abovedisplayskip}{3pt}
\setlength{\belowdisplayskip}{3pt}
\textbf{minimize} \max({\max} \{P(x'*r)_i: i \neq t\} - P(x'*r)_t, -\kappa),
\label{eq:optim}
\end{equation}
where $\{P(x'*r)_i: i \neq t\}$ represents the output probabilities of all speakers except the targeted speaker, while $P(x'*r)_t$ denotes the predicted probability to the targeted speaker. $\kappa$ is a configurable parameter which represents attack confidence and is set to $0$ in our implementation. 
To generate the universal perturbation, we iteratively modify the trainable sequence, $\bigtriangleup \delta$, which is used for constructing $\delta$, with the entire training dataset until satisfying the desired attack success rate. 
For each training utterance, if the predicted probability of the targeted class is larger than other classes, the update of the perturbation $\bigtriangleup \delta$ is skipped on the next sample.

\vspace{-5mm}
\subsection{Room Impulse Response Estimation}
\label{sec:rir}
\vspace{-2mm}
Acoustic propagation in a room is commonly considered as a linear and time-invariant system. Thus the recorded signal $R(x)$ could be presented as a deterministic function of the played signal $x$: $R(x)$ = $x*r$, where $r$ is the estimated room impulse response (RIR), and $*$ denotes the convolution operation. To simulate the play-over-the-air process in the physical world, we take the RIR generated by an acoustic room simulator~\cite{scheibler2018pyroomacoustics} into account in the adversarial example training phase. Specifically, the simulator
can adjust several parameters, including the size of a 3D shoe-box room, the location of the audio sources and microphones, and the reverberation rate. Optimization with the simulated RIR would increase the robustness of the generated adversarial example, and consequently enable over-the-air attack in practice.


\vspace{-4mm}
\section{Experimental Results}
\label{sec:evaluation}
\vspace{-3mm}

\begin{table}[t]
\centering
\caption{Results of universal targeted attack.}
\scalebox{0.8}{
\begin{tabular}{|c|c|c|c|c|}
\hline
\begin{tabular}[c]{@{}c@{}}Attack\\Strength \end{tabular} & \begin{tabular}[c]{@{}c@{}}Noise \\ Level \end{tabular} & \begin{tabular}[c]{@{}c@{}}Min. Attack\\Success Rate\end{tabular} & \begin{tabular}[c]{@{}c@{}}Max. Attack\\Success Rate\end{tabular}  &  \begin{tabular}[c]{@{}c@{}}Avg. Attack\\Success Rate\end{tabular}  \\ \hline
$\epsilon$=0.05 & -18.84dB& 98.47\% & 100\%   & 99.95\%      \\ \hline
$\epsilon$=0.03 & -23.27dB& 95.31\% & 99.91\%   &  98.40\%     \\ \hline
$\epsilon$=0.01 & -33.96dB & 53.32\%   & 95.48\%   & 83.82\%         \\ \hline
\end{tabular}}
\label{tab:dig}
\vspace{-5mm}
\end{table}

\subsection{Experimental Methodology}
\vspace{-2mm}

\textbf{Dataset}. We evaluate our proposed attack on an English multi-speaker corpus provided in CSTR voice cloning toolkit (VCTK)~\cite{christophe2016cstr}. In total, the dataset contains $44217$ utterances spoken by $109$ speakers with various accents. The
dataset is divided into a training and a testing set with a ratio of 4:1.

\noindent \textbf{Baseline Model}. 
In our TensorFlow-implemented X-vector system~\cite{snyder2018x}, 30-dimensional MFCC features with a frame length of $25ms$ are extracted. A pre-trained X-vector DNN embedding model provided in Kaldi~\cite{povey2011kaldi} is used in the model.
The baseline model achieves a classification accuracy of $92.8\%$ on $8896$ testing utterances from $109$ speakers. 

\noindent \textbf{Evaluation Metrics}. \textit{(1) Attack Success Rate}: The ratio between the number of succeeded attacks and the total number of attack attempts; \textit{(2) Noise Level}: We quantify the relative noise level of the perturbation $\delta$ with respect to the original audio $x$ in decibels (dB): $D(\delta ,x)=20log_{10}( \frac{max(\delta)}{max(x)})$.

\vspace{-5mm}
\subsection{Attack Evaluation}
\label{sec:attack_evaluation}
\vspace{-2mm}


\noindent \textbf{Effectiveness of Universal Targeted Attack}.
To evaluate the effectiveness of our proposed universal targeted attack, we alternatively choose one of the $109$ enrolled speakers as the targeted speaker and the rest $108$ speakers as victims. In total, we generated $109$ universal adversarial perturbations, trying to make the speaker recognition system classify the victims' utterances as the targeted speakers.
As shown in Table~\ref{tab:dig}, by adjusting attack strength $\epsilon$, the noise level ranges from $-18.84dB$ to $-33.96dB$. As discussed in the previous study~\cite{carlini2018audio}, such noise level is considered to be quasi-imperceptible to humans. For instance, $-33.96dB$ is comparatively the difference between a person talking and the ambient noise in a quiet room.
For each $\epsilon$ value, the minimum, maximum, and average attack success rate among all attack attempts targeting on $109$ speakers are calculated. We can observe that when the noise level is $-18.84dB$, a high average attack success rate of $99.95\%$ can be reached. When the noise level decreases to $-33.96$ dB, the average attack success rate still remains over $80\%$, which illustrates the effectiveness of our proposed universal targeted attack.

\noindent \textbf{Robustness Analysis Using Room Simulator}.
An acoustic room simulator toolkit~\cite{scheibler2018pyroomacoustics} is used to simulate the audio propagation in a room environment. 
Specifically, a modeled room with a size of $5m\times5m\times3m$ is used, and $120$ locations of the loudspeaker and the microphone are chosen randomly in the room for RIR estimation.
For the estimated RIRs, $100$ locations are used to build the universal, targeted and robust adversarial perturbation, and the rest $20$ locations are used for testing. 
Table~\ref{tab:rir} summarizes the results of our practical universal adversarial perturbation. We can observe that the universal adversarial perturbations trained with RIRs still remain effective after the over-the-air simulation. In particular, the practical universal perturbation generated with a noise level of $-18.84dB$ can still achieve an average attack success rate of $90.19\%$.
For comparison, we test the adversarial perturbation of the same noise level and without RIR in the simulated room environment. 
However, the average attack success rate decreases significantly to $1.33\%$. This shows that our approach can efficiently improve the robustness of the generated adversarial examples.

\noindent \textbf{Speedup on Attack Time}.
Unlike conventional individual attacks that require to build adversarial perturbation for each individual voice input, our proposed universal attack could generate a single perturbation that makes arbitrary speaker's utterances to be identified as the adversary-desired speaker. Thus, simply playing the pre-generated universal perturbation nearby the victim speaker becomes possible for launching adversarial attacks.
For showing the possibility of launching real-time attacks, we 
compare the attack launching time of using the conventional individual targeted attack method~\cite{carlini2018audio} and our proposed universal attack for a given audio signal. Particularly, the conventional targeted attack requires at least $15s$ to deploy, measured on a Tesla V100 GPU with $32GB$ memory, while our proposed universal method only takes an average of $0.015s$, which results in a \textit{$100\times$} speedup.

\begin{table}
\centering
\caption{Results of robust universal targeted attack using acoustic room simulator.}
\scalebox{0.7}{
\begin{tabular}{|c|c|c|c|c|}
\hline
& Noise Level & \begin{tabular}[c]{@{}c@{}}Min. Attack\\Success Rate\end{tabular} & \begin{tabular}[c]{@{}c@{}}Max. Attack\\Success Rate\end{tabular} & \begin{tabular}[c]{@{}c@{}}Avg.Attack\\Success Rate\end{tabular} \\ \hline
Without RIR  & -18.84dB    & 0.7\%   & 3.52\%  & 1.33\%  \\ \hline

& -18.84dB    & 74.68\% & 98.05\% & 90.19\% \\ \cline{2-5} 
\multirow{2}{*}{With RIR}  & -23.27dB    & 66.54\% & 96.81\% & 86.17\% \\ \cline{2-5} 
& -33.96dB    & 54.48\% & 90.83\% & 78.25\% \\  \hline
\end{tabular}}
\label{tab:rir}
\vspace{-4mm}
\end{table}

\vspace{-4mm}
\section{Conclusion}
\label{sec:conclusion}
\vspace{-3mm}
This paper proposes a real-time, universal, and robust targeted adversarial attack against speaker recognition system. The proposed attack builds a universal perturbation that can be added into any enrolled speaker's voice input to fool the system causing it to output any adversary-desired speaker label. The robustness of the adversarial perturbations is also greatly improved by using an acoustic room simulator to estimate the sound distortions associated with playing the audio over-the-air. Evaluation on a public dataset of $109$ speakers shows the effectiveness and robustness of our proposed attack.

\noindent\textbf{Acknowledgments}
This research is supported in part by the National Science Foundation grants CNS1801630 and CCF1909963, the Army Research Office grant W911NF-18-1-0221
and the Air Force Research Laboratory grant FA8750-18-2-0058.


\clearpage
\bibliographystyle{IEEEbib}
\bibliography{refs}

\begin{thebibliography}{10}

\bibitem{voice_match}
Google,
\newblock ``Voice match and media on google home,''
  \url{https://support.google.com/googlenest/answer/7342711?hl=en}, Sep. 2019.

\bibitem{chase}
Chase Bank,
\newblock ``Security as unique as your voice,''
  \url{https://www.chase.com/personal/voice-biometrics}, Oct. 2019.

\bibitem{lei2014novel}
Yun Lei, Nicolas Scheffer, Luciana Ferrer, and Mitchell McLaren,
\newblock ``A novel scheme for speaker recognition using a phonetically-aware
  deep neural network,''
\newblock in {\em 2014 IEEE International Conference on Acoustics, Speech and
  Signal Processing (ICASSP)}. IEEE, 2014, pp. 1695--1699.

\bibitem{mclaren2015advances}
Mitchell McLaren, Yun Lei, and Luciana Ferrer,
\newblock ``Advances in deep neural network approaches to speaker
  recognition,''
\newblock in {\em 2015 IEEE international conference on acoustics, speech and
  signal processing (ICASSP)}. IEEE, 2015, pp. 4814--4818.

\bibitem{intriguing}
Christian Szegedy, Wojciech Zaremba, Ilya Sutskever, Joan Bruna, Dumitru Erhan,
  Ian Goodfellow, and Rob Fergus,
\newblock ``Intriguing properties of neural networks,''
\newblock {\em arXiv preprint arXiv:1312.6199}, 2013.

\bibitem{carlini2018audio}
Nicholas Carlini and David Wagner,
\newblock ``Audio adversarial examples: Targeted attacks on speech-to-text,''
\newblock in {\em 2018 IEEE Security and Privacy Workshops (SPW)}. IEEE, 2018,
  pp. 1--7.

\bibitem{carlini2016hidden}
Nicholas Carlini, Pratyush Mishra, Tavish Vaidya, Yuankai Zhang, Micah Sherr,
  Clay Shields, David Wagner, and Wenchao Zhou,
\newblock ``Hidden voice commands,''
\newblock in {\em 25th $\{$USENIX$\}$ Security Symposium ($\{$USENIX$\}$
  Security 16)}, 2016, pp. 513--530.

\bibitem{kreuk2018fooling}
Felix Kreuk, Yossi Adi, Moustapha Cisse, and Joseph Keshet,
\newblock ``Fooling end-to-end speaker verification with adversarial
  examples,''
\newblock in {\em 2018 IEEE International Conference on Acoustics, Speech and
  Signal Processing (ICASSP)}. IEEE, 2018, pp. 1962--1966.

\bibitem{snyder2018x}
David Snyder, Daniel Garcia-Romero, Gregory Sell, Daniel Povey, and Sanjeev
  Khudanpur,
\newblock ``X-vectors: Robust dnn embeddings for speaker recognition,''
\newblock in {\em 2018 IEEE International Conference on Acoustics, Speech and
  Signal Processing (ICASSP)}. IEEE, 2018, pp. 5329--5333.

\bibitem{vaidya2015cocaine}
Tavish Vaidya, Yuankai Zhang, Micah Sherr, and Clay Shields,
\newblock ``Cocaine noodles: exploiting the gap between human and machine
  speech recognition,''
\newblock in {\em 9th $\{$USENIX$\}$ Workshop on Offensive Technologies
  ($\{$WOOT$\}$ 15)}, 2015.

\bibitem{yuan2018commandersong}
Xuejing Yuan, Yuxuan Chen, Yue Zhao, Yunhui Long, Xiaokang Liu, Kai Chen,
  Shengzhi Zhang, Heqing Huang, XiaoFeng Wang, and Carl~A Gunter,
\newblock ``Commandersong: A systematic approach for practical adversarial
  voice recognition,''
\newblock in {\em 27th $\{$USENIX$\}$ Security Symposium ($\{$USENIX$\}$
  Security 18)}, 2018, pp. 49--64.

\bibitem{neekhara2019universal}
Paarth Neekhara, Shehzeen Hussain, Prakhar Pandey, Shlomo Dubnov, Julian
  McAuley, and Farinaz Koushanfar,
\newblock ``Universal adversarial perturbations for speech recognition
  systems,''
\newblock {\em arXiv preprint arXiv:1905.03828}, 2019.

\bibitem{snyder2019speaker}
David Snyder, Daniel Garcia-Romero, Gregory Sell, Alan McCree, Daniel Povey,
  and Sanjeev Khudanpur,
\newblock ``Speaker recognition for multi-speaker conversations using
  x-vectors,''
\newblock in {\em ICASSP 2019-2019 IEEE International Conference on Acoustics,
  Speech and Signal Processing (ICASSP)}. IEEE, 2019, pp. 5796--5800.

\bibitem{raj2019probing}
Desh Raj, David Snyder, Daniel Povey, and Sanjeev Khudanpur,
\newblock ``Probing the information encoded in x-vectors,''
\newblock {\em arXiv preprint arXiv:1909.06351}, 2019.

\bibitem{snyder2017deep}
David Snyder, Daniel Garcia-Romero, Daniel Povey, and Sanjeev Khudanpur,
\newblock ``Deep neural network embeddings for text-independent speaker
  verification,''
\newblock in {\em Interspeech}, 2017, pp. 999--1003.

\bibitem{scheibler2018pyroomacoustics}
Robin Scheibler, Eric Bezzam, and Ivan Dokmani{\'c},
\newblock ``Pyroomacoustics: A python package for audio room simulation and
  array processing algorithms,''
\newblock in {\em 2018 IEEE International Conference on Acoustics, Speech and
  Signal Processing (ICASSP)}. IEEE, 2018, pp. 351--355.

\bibitem{christophe2016cstr}
Veaux Christophe, Yarnagishi Junichi, and MacDonald Kirsten,
\newblock ``Cstr vctk corpus: English multi-speaker corpus for cstr voice
  cloning toolkit,''
\newblock {\em The Centre for Speech Technology Research (CSTR)}, 2016.

\bibitem{povey2011kaldi}
Daniel Povey, Arnab Ghoshal, Gilles Boulianne, Lukas Burget, Ondrej Glembek,
  Nagendra Goel, Mirko Hannemann, Petr Motlicek, Yanmin Qian, Petr Schwarz,
  et~al.,
\newblock ``The kaldi speech recognition toolkit,''
\newblock in {\em IEEE 2011 workshop on automatic speech recognition and
  understanding}. IEEE Signal Processing Society, 2011, number CONF.

\end{thebibliography}

\end{document}